\documentclass[10pt,letterpaper]{article}
\usepackage[top=0.85in,left=2.75in,footskip=0.75in]{geometry}
\RequirePackage{fix-cm}
\usepackage{graphicx}
\usepackage{lipsum}
\usepackage{ulem}
\usepackage{xcolor}
\usepackage{color}
\usepackage{nicefrac}
\usepackage{algorithmic}
\usepackage{tabularx}
\usepackage{longtable}
\usepackage{empheq}
\usepackage{booktabs}
\usepackage{multirow}
\usepackage[linesnumbered,ruled,resetcount]{algorithm2e}
\usepackage{algorithmic}
\usepackage{enumitem}
\usepackage{setspace}
\usepackage{tabu}
\usepackage{cancel}
\usepackage{optidef}
\usepackage{subcaption}
\usepackage[hyphens]{url}

\usepackage{amsmath,amssymb}

\usepackage{changepage}

\usepackage[utf8x]{inputenc}

\usepackage{textcomp,marvosym}

\usepackage{cite}

\usepackage[right]{lineno}

\usepackage{microtype}
\DisableLigatures[f]{encoding = *, family = * }

\usepackage{array}

\newcolumntype{+}{!{\vrule width 2pt}}

\newlength\savedwidth

\newcommand\thickhline{\noalign{\global\savedwidth\arrayrulewidth\global\arrayrulewidth 2pt}%
\hline
\noalign{\global\arrayrulewidth\savedwidth}}

\newcount\colveccount
\newcommand*\colvec[1]{
        \global\colveccount#1
        \begin{bmatrix}
        \colvecnext
}
\newcommand{\colvecnext}[1]{
		#1
        \global\advance\colveccount-1
        \ifnum\colveccount>0
                \\
                \expandafter\colvecnext
        \else
                \end{bmatrix}
        \fi
}

\usepackage[aboveskip=1pt,labelfont=bf,labelsep=period,justification=raggedright,singlelinecheck=off]{caption}

\makeatletter
\renewcommand{\@biblabel}[1]{\quad#1.}
\makeatother

\usepackage{lastpage,fancyhdr,graphicx}
\usepackage{epstopdf}
\fancyheadoffset[L]{2.25in}
\fancyfootoffset[L]{2.25in}
\begin{document}
\vspace*{0.2in}

\begin{flushleft}
{\Large
\textbf\newline{Game theory to enhance stock management of personal protective equipment (PPE) during the COVID-19 outbreak} 
}
\newline
\\
Khaled Abedrabboh\textsuperscript{1},
Matthias Pilz\textsuperscript{2},
Zaid Al-Fagih\textsuperscript{3},
Othman S. Al-Fagih\textsuperscript{4},
Jean-Christophe Nebel\textsuperscript{5},
Luluwah Al-Fagih\textsuperscript{5,6*}
\\
\bigskip
\textbf{1} Division of Sustainable Development, College of Science and Engineering, Hamad Bin Khalifa University, Qatar Foundation, Doha, Qatar
\\
\textbf{2} Independent Researcher, e-mail: Pilz.Matthias@outlook.com
\\
\textbf{3} Independent Researcher, e-mail: zaid.al-fagih@nhs.net
\\
\textbf{4} NHS Health Education East of England, Cambridge, UK
\\
\textbf{5} School of Computer Science and Mathematics, Kingston University, London, UK
\\
\textbf{6} Division of Engineering Management and Decision Sciences, College of Science and Engineering, Hamad Bin Khalifa University, Qatar Foundation, Doha, Qatar
\\
\bigskip

%
%





*Corresponding author. Email: lalfagih@hbku.edu.qa (LAF)

\end{flushleft}
\section*{Abstract}

Since the outbreak of the COVID-19 pandemic, many healthcare facilities have suffered from shortages in medical resources, particularly in Personal Protective Equipment (PPE). In this paper, we propose a game-theoretic approach to schedule PPE orders among healthcare facilities. In this PPE game, each independent healthcare facility optimises its own storage utilisation in order to keep its PPE cost at a minimum. Such a model can reduce peak demand considerably when applied to a variable PPE consumption profile. Experiments conducted for NHS England regions using actual data confirm that the challenge of securing PPE supply during disasters such as COVID-19 can be eased if proper stock management procedures are adopted. These procedures can include early stockpiling, increasing storage capacities and implementing measures that can prolong the time period between successive infection waves, such as social distancing measures. Simulation results suggest that the provision of PPE dedicated storage space can be a viable solution to avoid straining PPE supply chains in case a second wave of COVID-19 infections occurs.


\section{Introduction}
\label{sec:intro}
Novel infectious diseases pose a serious challenge to policy makers and healthcare systems. Emerging from Wuhan, China, the ongoing Coronavirus Disease 2019 (COVID-19) pandemic, caused by the Severe acute respiratory syndrome coronavirus 2 (SARS-CoV-2), has wreaked havoc globally. This is due to its rapid rate of transmission, its virulence and the inability of most countries to adequately prepare for such a disease \cite{Madabhavi2020}.  
Identified in December 2019, the disease now has a global distribution with over 30 million confirmed cases and almost one million confirmed deaths as of September 2020 according to the World Health Organisation (WHO) \cite{WorldHealthOrganizationWHO2020c}. 
COVID-19 is primarily transmitted through respiratory droplets and the WHO has identified two principal routes through which these are carried between people. The first mode of transmission involves a person being in direct, close contact with someone who carries the virus (within one metre) where they become directly exposed to potentially infectious respiratory droplets. The second mode of transmission involves contact with fomites  in the immediate vicinity of the infected person \cite{Ong2020}.
Nguyen \textit{et al.}~estimated that frontline healthcare professionals (HCPs) had a 3.4 times higher risk than non-healthcare workers of contracting COVID-19, even when adjusting for the probability of being tested  \cite{Nguyen2020}. Indeed, approximately 10\% of the confirmed cases in China \cite{WorldHealthOrganizationWHO2020b} and up to 9\% of all cases in Italy have been among healthcare professionals \cite{IstitutoSuperiorediSanitaISS2020} as of the date of publication of these studies. This increased risk does not only pose a problem for the HCPs themselves, but also poses a major threat to the elderly and vulnerable populations they care for, since outbreaks within healthcare settings are important amplifiers of infection \cite{EuropeanCentreforDiseasePreventionandControlECDC2020}. One of the most crucial ways by which transmission of this virus (and other infectious diseases) is reduced is the use of proper Personal Protective Equipment (PPE) \cite{EuropeanCentreforDiseasePreventionandControlECDC2020}: 
the WHO recommends a surgical mask, goggles, or face shield, gown, and gloves to be worn as PPE in their COVID-19 PPE guidelines. If an aerosol-generating procedure is performed, the surgical mask is replaced with an N95 or FFP2/3 respirator which provides a greater level of filtration than surgical masks  \cite{WorldHealthOrganizationWHO2020d}. These guidelines are replicated globally with minor differences. In the United Kingdom (the case study reported in this paper), the most recent Public Health England guidelines are essentially the same \cite{PublicHealthEnglandPHE2020a}. 
Globally, medical resources, particularly PPE have come under unprecedented demand. This has led to shortages in many countries with some rationing their use of PPE and in some cases reusing disposable material \cite{Tabah2020}. The natural consequence of this has been the disproportionately high rate of infection amongst HCPs which in turn contributes to disease spread. This may not necessarily result from a national shortage of PPE but rather local shortages resulting from inefficient distribution of resources in timely manner \cite{Reuters2020}. 

In spite of the limited applicability of resource allocation methods, as choices have to be made, they can aid in making the decisions that achieve the best health outcomes (see\cite{Brandeau2005} for a review on the use of such methods in epidemic control). Single- and multi-objective optimisation methods have been used frequently to address problems  of resource allocation and scheduling of purchase orders for medical supplies (see \cite{tuzun2018taxonomy} for a comprehensive survey and taxonomy). Such methods are centralised in the sense that a central planner seeks to optimally coordinate supply activities for the entire system, e.g.~minimise overall costs for the central planner. In contrast to this, game theory is a tool that allows decentralised decision-making~\cite{ShohamBrown2008,Neumann}. That means, different entities in the system can make decisions based on their individual preferences. For instance, they can schedule their orders to minimise their \textit{own} costs. The decentralised approach leaves the actors with more freedom and is the more applicable direction for our scenario.

Game theory has been applied to a variety of subject areas such as biology \cite{Hammerstein1994}, economics \cite{Cournot1838}, computer science \cite{Shoham2008} and energy \cite{Pilz2017IEEESmartGrid,Pilz2019Cyber}.
In general, game theory is used to mathematically model systems of competing agents. Usually these individuals act in a selfish and rational manner. In this context, \textit{selfish} can be understood as being only interested in their own good, i.e.~an agent strives to maximise its outcome irrespective of the outcome of others~\cite{ShohamBrown2008}. Furthermore, \textit{rational} means that there is a clear logical reasoning behind every decision~\cite{Neumann}. The most widely used solution concept for a non-cooperative game is the Nash equilibrium \cite{Nash1951}. It is achieved when none of the players has an incentive to change their strategy unilaterally. 

Despite its various applications in supply chain management (cf.~\cite{Vasnani2019} and the references therein), only few studies have applied game theory to the management of medical supplies. To the best of our knowledge, a 2008 study \cite{DeLaurentis2008} (an extended version was later published in \cite{Adida2011}) was the first to develop a game theoretic approach for stockpiling of critical medical items. In their model, the authors propose a non-cooperative game where hospitals stockpile critical medical resources in preparation for disasters. It is proposed that these hospitals determine their individual stockpile levels strategically in a way that minimises their expected total spend. Their model uses a cost function that consists of the cost of ordering, borrowing, storage and a penalty cost in the case of shortages. Although their model proves to give some breathing space to the medical supply chain at the onset of an epidemic, it is only intended as a preparation scheme, and therefore may suffer from inapplicability if an epidemic lasts longer than the planning period considered. The model assumes a given likelihood of the occurrence and severity of a pandemic, which is, in practice, heavily unpredictable. Their work was further developed in \cite{Lofgren2016} by introducing network constraints, thus giving realistic hospital sharing policies. The authors find that deficits between stockpiles and demand can be reduced through central stockpiling and through increasing penalties for deficits.

Game theory has also been applied to the problem of drug allocation \cite{Sun2009}, where countries behave selfishly to minimise their expected number of infections. Unlike in \cite{Wang2009} where the authors propose a similar resource competition among countries, their model imitates the stochasticity of infection transmission parameters. The resulting comparison between this selfish allocation scheme and a utilitarian division scheme, where a central planner (e.g.~WHO) makes all of the allocation decisions, shows that having a central planner reduces the total number of worldwide infections considerably. Although their proposed model can aid in understanding how countries, acting in their own self interest, would behave in an epidemic, it does not address the problem of resource distribution within a given country. \cite{Liu2012} compares selfish vaccination coverage, which the authors call the `Nash vaccination`, with group optimal coverage, which they call the `utilitarian vaccination'. The authors find that the cost of vaccination would be a pivotal factor in determining the effectiveness of Nash vs.~utilitarian coverage. However, this can result in different outcomes depending on the level of disease severity according to age, such as in the case of chickenpox. 

Most recently, Nagurney \textit{et al.}~proposed a novel decentralised model for medical supply chain with multiple supply and demand points  \cite{NagurneyAnna;SalarpourMojtaba;DongJune;Dutta2020}. In their model, selfish consumers who have stochastic demands make purchasing decisions that minimise their total expenditure. The authors assume that the disutility function of a consumer consists of a linear cost of demand, a quadratic cost of transportation and a penalty for shortages/surpluses. Although the quantification of the shortage/surplus penalty can be debated, the authors conclude by suggesting that shortages in supply can be avoided by redirecting global production efforts to instead adding or increasing local production capabilities.

In this paper, we propose a game theoretic approach for managing PPE supplies during a pandemic. We take inspiration from the electricity storage scheduling game developed by the co-authors in \cite{pilz2019dynamic}, where a  decentralised system of individually owned home energy systems served by the same utility company schedule their day-ahead battery usage over a full year.  This decentralised system resulted in improved energy efficiency for the utility company and cost savings for the participating users. 

This work proposes a \textit{centralised-decentralised} approach to the PPE supply chain (cf.~Section~\ref{sec:mat}). In the proposed architecture, healthcare facilities report their demand to a central entity. This central entity is assumed to have the commitment power to fulfill its orders and set the costs for PPE. Given the actions of all game participants, healthcare facilities optimise their PPE orders by making stockpiling decisions individually. Our approach is centralised in the sense that cost and supply is controlled by a central entity. It is also decentralised because healthcare facilities make their stockpiling decisions independently. By adapting the model developed in \cite{pilz2019dynamic} and applying it to COVID-19 related PPE demand in England, we study the effects of early stockpiling as well as increasing storage capacities on PPE supply in challenging circumstances. Additionally, we examine the impact of a putative second wave on PPE supply and investigate whether delaying this putative second wave can ease the challenge of fulfilling the PPE demand of healthcare facilities. Finally, insights and conclusions from this study are drawn and suggestions regarding PPE supply in disaster management are discussed. The contributions of this work can be summarised as follows:
\begin{enumerate}[itemsep=0pt, leftmargin=1cm, labelsep=0.3cm]
    \item A game theory-based model designed to enhance stock management of PPE supply.
	\item A study of PPE supply management in England during the current COVID-19 pandemic demonstrating the benefits of the centralised-decentralised allocation approach advocated by the proposed model.
	\item A detailed analysis of how key factors, i.e. stockpiling start date, storage capacity and the date of a putative second wave, impact PPE supply in England.
	\item Insights and suggestions regarding the handling of a putative second wave in terms of PPE supply management.
\end{enumerate}

The remainder of this paper is organised as follows. In Section~\ref{sec:mat} we provide an overview of the system architecture, explain the chosen cost function and give details on the game formulation. Section~\ref{sec:res} contains information on the experimental setup and Section~\ref{sec:res2} presents an analysis of the results with simulations of different scenarios. This is followed by a discussion and recommendations in Section~\ref{sec:dis} and finally, Section~\ref{sec:con} concludes the paper and points out future research directions. 

\section{System model}
\label{sec:mat}

During the COVID-19 pandemic, most countries have experienced strained healthcare resources and even shortages. The efficient management of personal protective equipment (PPE) supply has proved vital in limiting the spread of the virus and in keeping the healthcare professionals protected.
In this paper, after designing a game theory-based model to enhance stock management of PPE supply, we investigate as a case study PPE provision in hospitals of the English National Health Service (NHS). England was selected as not only has it been one of the first and most severely hit countries, but it has also released COVID-19 data in a transparent and timely manner. The datasets and assumptions that were used in this case study are described in Section~\ref{sec:res}. Data on occupied hospital beds with illness related to COVID-19 \cite{PublicHealthEnglandPHE2020} was used as a proxy to generate PPE demand profiles.

\subsection*{System architecture}

Healthcare facilities in England are run by the NHS (NHS England). The NHS is made up of organisational units named Trusts which mainly serve geographic areas, but can also serve specialised functions. For a detailed description of the NHS structure in England, please see \cite{NHS2014}. Prior to the outbreak of COVID-19 in England, procurement of medical supplies to NHS Trusts, including PPE, was centralised where each Trust would ‘order’ supplies via NHS Supply Chain \cite{SCCL}. However, in the initial period of the pandemic, the NHS was unable to fulfil Trusts' demand centrally leading to a chaotic albeit temporary decentralised supply chain; some of which was unconventional \cite{Milmo2020}. On 01 May, NHS Supply Chain introduced a dedicated PPE supplies channel separate to other medical supplies to address this issue \cite{DepartmentofHealth&SocialCareDHSC}.

\begin{figure}[h]
\centering
\includegraphics[width=0.8\textwidth]{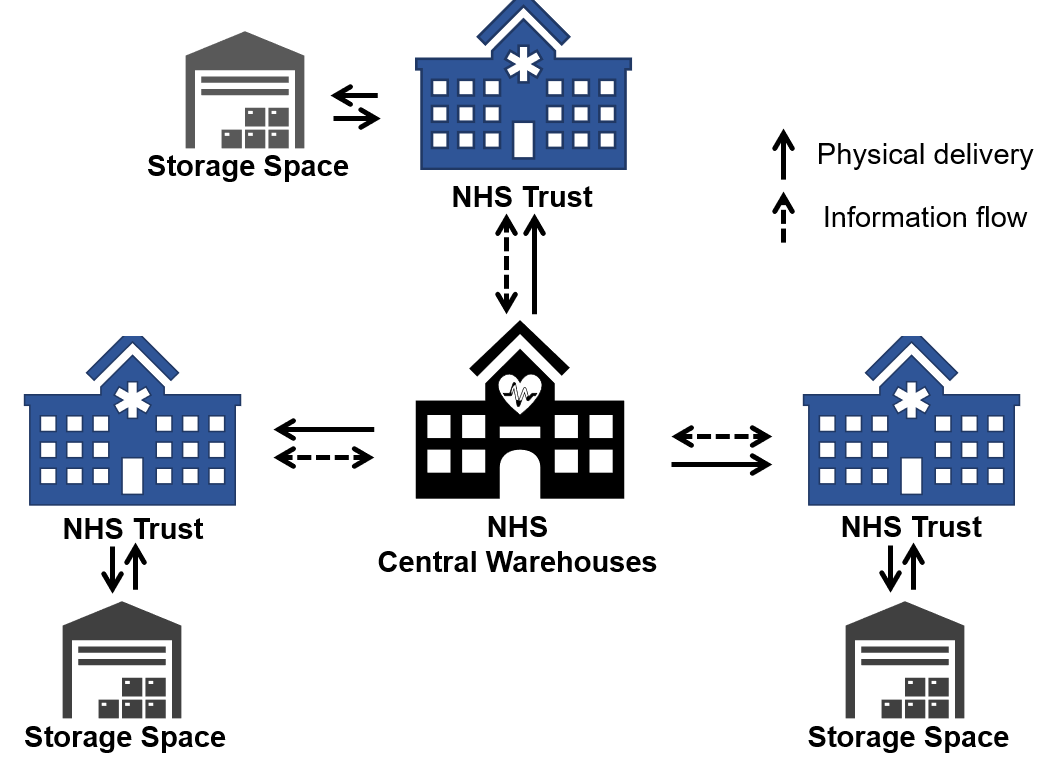}
\caption{\textbf{System architecture for the proposed model. It shows the centralised-decentralised approach to PPE supply chain.}}
\label{fig:Fig 1}
\end{figure}

The NHS openly expresses its commitment to improve efficiency \cite{NHSImprovementa} and, via its NHS Improvement department, it has advocated the use of modelling and novel ideas to facilitate this \cite{NHSImprovementb}. In line with favoured practice by the NHS, we propose a centralised-decentralised PPE supply chain architecture, shown in Fig~\ref{fig:Fig 1}, where a central entity (controlled by the NHS) defines the pricing function based on the market price of the sourced PPE. Thus, for a given day, all NHS Trusts will pay the same price per PPE item.

Similar to the decentralised healthcare resource allocation game proposed in \cite{Wang2009}, in this architecture, each NHS Trust is modelled as an independent entity that is selfishly concerned with their own interests of satisfying their demand and minimising their cost. Thus, NHS Trusts individually manage the use of their storage capabilities so that their PPE cost is minimised. After optimising their storage schedules independently, NHS Trusts then report their optimal PPE orders to the central entity, which in turn fulfils their demand and charges them their share of PPE cost. An important feature of our model is that the daily PPE demand is always fulfilled. Therefore, patients are always prioritised over costs and thus the stock management strategy does not have any public health implications.

\subsection*{Cost function}

In order to ensure security of supply and avoid burdening the supply chain, sudden surges in demand should be eliminated or at least planned for. It is therefore beneficial for the central entity to aim at having a relatively flat demand profile. This can be done by incentivising Trusts to utilise their storage in an optimal manner so that their overall demand is level. Accordingly, the following assumptions are made regarding the cost function of PPE, $C(Q)$, where $Q$ is the total quantity of PPE consumption:
\begin{enumerate}[itemsep=0pt, leftmargin=1cm, labelsep=0.3cm]
	\item The cost function of PPE $C(Q)$ is \textit{strictly increasing} in consumption $Q$,
	\begin{equation}
			\frac{\mathrm{d} C}{\mathrm{d} Q}>0, 
		\label{eqn:cost_assumption}
	\end{equation}
		i.e.~the higher the consumption the higher the cost.
	\item The marginal cost of PPE is \textit{strictly increasing} in consumption, 
	\begin{equation}
			\frac{\mathrm{d}^2 C}{\mathrm{d} Q^2}>0,  
		\label{eqn:cost_assumption2}
	\end{equation}
		i.e.~the rate of rise of cost increases when consumption increases. This assumption can be justified by considering that an increase in demand will likely require new supply routes or new production facilities be established.
	\item Zero consumption yields (incurs) zero cost
	\begin{equation}
			C\left( 0 \right ) = 0.  
		\label{eqn:cost_assumption3}
	\end{equation}
\end{enumerate}

While many functions satisfy the above assumptions, this work adopts a quadratic cost function where
\begin{equation}
	C(Q)= aQ^2 + bQ  
	\label{eqn:quadratic_cost}
\end{equation}
and where $a>0,~b\geq0$ are constant cost parameters, 
as is the case in \cite{NagurneyAnna;SalarpourMojtaba;DongJune;Dutta2020}, where the authors use a quadratic cost function for transportation of PPE along with a linear cost function for PPE supply.   
The use of quadratic cost function of PPE is further supported by the documented rise in PPE costs during peak COVID-19 cases in England. In some instances, a 1000\% increase in prices was reported by NHS Trusts \cite{Hardy2020}.

\subsection*{Game formulation}

The notations used in this formulation are defined in Table~\ref{Table 1}. Within our model the consumption $Q$ (as introduced in the previous section) consists of two separate quantities. On the one hand, there is the demand $d$ for PPE according to the number of patients that are currently treated. This number cannot directly be influenced within the game formulation. On the other hand, there is the number of PPE kits $a$ that are put into the storage (or taken from the storage), which is our decision to make. Thus we have: 
\begin{equation}
    Q = d + a\ .
\end{equation} While $d$ will always be larger than zero, $a$ can take values in a range from $max(-d, -s)$ to $s_{\text{max}}-s$, where $s$ is the amount of PPE currently available in storage and $s_{\text{max}}$ is the storage capacity. Thus, the largest amount of PPE that can be taken from storage is limited by either the current demand or the amount of available stored PPE, whichever is lower. Additionally, the maximum amount of PPE that can be stored is bounded by the storage capacity.
Formally, this can be summarised by the following constraint: 
\begin{equation}
    \mathbf{h}(s,a) = \colvec{2}{a - (s_{\text{max}}-s)}{-a + \text{min}(d, s)}\ .
    \label{eqn:game_constraint_h}
\end{equation}

The chosen action $a$ then directly affects the stored PPE leading to the following transition equation:
\begin{equation}
    f(s_{\text{old}},a) = s_{\text{new}} = s_{\text{old}} + a \ .
    \label{eqn:game_transition}
\end{equation}

\setcounter{table}{0}
\begin{table}[!ht]
\setlength{\extrarowheight}{0.15cm}
\centering
\caption{\textbf{Notations and associated definitions.}}
\label{Table 1}
\begin{tabular}{>{\centering\arraybackslash}m{2cm}|m{10.4cm}}

\multicolumn{1}{c}{Notation} \vline & 
\multicolumn{1}{c}{Definition}\\
\thickhline
$\mathcal{N}$       & Set of players\\ 
$N$                 & Number of players\\ 
$n$                 & Index of a player, where $n \in \mathcal{N}$\\ 
$\mathcal{T}$       & Set of intervals\\ 
$T$                 & Number of intervals\\ 
$t$                 & Index of an interval, where $t \in \mathcal{T}$\\ 
$d_n^t$             & PPE demand of player $n$ at interval $t$\\ 
$s_n^t$             & Amount of stockpiled PPE (state of storage) of player $n$ at interval $t$\\ 
$a_n^t$             & Amount of PPE to be stored/used by player $n$ at interval $t$, this denotes the decision to either add to the storage ($+ve~a$) or take from it ($-ve~a$)\\ 
$\mathcal{H}_n^t$   & Set of admissible decisions for player $n$ at interval $t$\\ 
$u_n$               & Utility of player $n$\\ 
$C_n^t$             & Cost of PPE for player $n$ at interval $t$\\ 
$C_n^T$             & Cost of PPE for player $n$ at end of game cycle $T$\\ 
$G_n$               & Decision problem (game) of player $n$\\ 
\end{tabular}
\end{table}

Similar to \cite{pilz2019dynamic}, we propose a discrete time dynamic game, where the decisions of the players (the Trusts) of how much to put/take in/from the PPE storage are performed sequentially in stages. These stages (also called intervals) are defined according to the actual demand variations, i.e.~if the demand changes on a daily basis, each interval would cover a one day period. Furthermore we introduce the state of the game (for each interval) and how it interacts with the decisions of the players. Overall the goal of the players is to minimise their own costs of PPE, i.e.~their utility function which is closely related to the cost function discussed in the previous section (see Eq~\ref{eqn:quadratic_cost}). To summarise, the game consists of:
{\setlength{\extrarowheight}{.5em}
	\begin{longtable}{p{.04\textwidth}p{.9\textwidth}}
	1. & A set of \textit{players} (Trusts) $\mathcal{N} = \{1,\dots,N\}$, where $N$ is the total number of participants. All players are assumed to be selfish and rational.\\
	2. & A set of \textit{intervals} (typically days), $\mathcal{T} = \{0,\dots,T-1\}$ where $T$ is the number of intervals that comprise the intended time cycle of the game.\\
	3. & Scalar state variables $s_n^t\in\mathcal{S}_n \subset{\mathrm{I\!R}}$ denoting the amount of stored PPE of the $n^{th}$ player at stage $t\in\mathcal{T}\cup\{T\}$. Collectively, we denote the state variables of all players at stage $t$ by $\mathbf{s}^t:=\left[s_1^t,\dots,s_N^t\right]\in\mathcal{S}:=\mathcal{S}_1\times\cdots\times\mathcal{S}_N\subset{\mathrm{I\!R}}^N$. In the open-loop information structure it is assumed that the \textit{initial state} $s^0$ is known to all players $n\in\mathcal{N}$.\\
	4. & Scalar \textit{decision variables} $a_n^t\in\mathcal{H}_n^t\left(s_n^t\right)\subset\mathcal{A}_n\subset{\mathrm{I\!R}}$ (for definition of $\mathcal{H}_n^t$ see item (5)) denoting the usage of the stored PPE of the $n^{th}$ player at time $t\in\mathcal{T}$. Collectively, we denote the decision variables of all players at stage $t$ by $\mathbf{a}^t:=\left[a_1^t,\dots,a_N^t\right]\in\mathcal{A}:=\mathcal{A}_1\times\cdots\times\mathcal{A}_N\subset{\mathrm{I\!R}}^N.$ Furthermore we define the \textit{schedule of PPE usage} of an individual player $n\in\mathcal{N}$ as a collection of all its decisions in the stages of the game by $\mathbf{a}_n:=\left[a_n^0,\dots,a_n^{T-1}\right]$. A \textit{strategy profile} is denoted by $\mathbf{a}:=\left[\mathbf{a}_1,\dots,\mathbf{a}_N\right]$.\\
	5. & A set of \textit{admissible decisions} $\mathcal{H}_n\left(s_n^0\right) := \left\{\mathbf{a}_n~|~h_n^t\left(s_n^t,a_n^t\right)\leq 0,\ t\in\mathcal{T}\right\}\subset{\mathrm{I\!R}}^T$ for the $n^{th}$ player. The function $h_n^t\left(s_n^t,a_n^t\right)$ has been defined in Eq~\ref{eqn:game_constraint_h}, capturing the restrictions posed by the storage facilities. We denote $\mathcal{H}_n^t\left(s_n^t\right):= \left\{a_n^t~|~h_n^t\left(s_n^t,a_n^t\right)\leq 0\right\}\subset{\mathrm{I\!R}}$\\
	6. & A \textit{state transition equation}
		\begin{equation}
			s_n^{t+1} = f_n^t\left(s_n^t, a_n^t\right),\ \ t\in\mathcal{T},\ n\in\mathcal{N},
			\label{eqn:SP_stateTransistion}
		\end{equation}
		governing the state variables $\left\{\mathbf{s}^t \right\}_{t=0}^T$. The function $f_n^t\left(s_n^t, a_n^t\right)$ is the discretised version of the transition equation (Eq~\ref{eqn:game_transition}), showing how a decision of the player influences the state of its PPE storage for the upcoming stage.\\
		7. & A \textit{stage additive utility function} 
			\begin{equation}
			    u_n\left(s_n^0, \left[\mathbf{a}_n, \mathbf{a}_{-n}\right]\right)=-C_n^T\left(s_n^T\right) - \sum_{t=0}^{T-1}C_n^t\left(s_n^t, \left[a_n^t, \mathbf{a}_{-n}^t\right]\right)
				\label{eqn:SP_additiveUtilityFunction}
			\end{equation}				
		 	for the $n$th player, where $\mathbf{a}_{-n}:=\left[\mathbf{a}_1,\dots,\mathbf{a}_{n-1},\mathbf{a}_{n+1},\dots,\mathbf{a}_N\right]$ denotes the decisions of all other players. The function $C_n^t\left(s_n^t, \left[a_n^t, \mathbf{a}_{-n}^t\right]\right)$ fulfils the assumptions as denoted in Eq~\ref{eqn:quadratic_cost} capturing the costs to the $n^{th}$ player at the $t^{th}$ stage. Note that the utility function depends only on the initial state variable $s_n^0$, since the subsequent states $s_n^t$ are determined by Eq~\ref{eqn:SP_stateTransistion}. The function 
						\begin{equation}
						    C_n^T\left(s_n^T\right) = s_n^T
							\label{eqn:SP_gT}
						\end{equation}				 
		 	can be interpreted as a penalty for the $n^{th}$ Trust that is incurred by ending up in state $s_n^T$, i.e.~its overbought PPE capacity, at the end of the scheduling period.
	\end{longtable}%
}

The objective of rational players is to maximise their total utility, i.e.~minimise their overall costs, over the complete time cycle $T$. We represent the decision problem $G_n$ of the $n^{th}$ player (given the actions $\mathbf{a}_{-n}$ of all the other players) as the following optimisation problem:
	\begin{empheq}{equation}
		\label{eqn:SP_dynamicGame}
		\begin{split}
  			G_n\left(\mathbf{a}_{-n}\right) \hspace{1cm}	& \makebox[0pt][l]{\text{given }}\phantom{\text{subject to }}\ \mathbf{s}^0\in\mathcal{S} \\
  				& \makebox[0pt][l]{$\underset{a_n}{\text{maximise }}$}\phantom{\text{subject to }}\ u_n\left(s_n^0,\left[\mathbf{a}_n, \mathbf{a}_{-n}\right]\right)\\
  				& \text{subject to }\ a_n^t\in\mathcal{H}_n^t\left(s_n^t\right)\\
  				& \hphantom{\text{subject to }}\ s_n^{t+1} = f_n^t\left(s_n^t, a_n^t\right) \ \forall t\in\mathcal{T}\cup\{T\}
		\end{split}
	\end{empheq}
	Moreover, the game is referred to as $\left\{ G_1,\dots,G_N\right\}$, which denotes the simultaneous (and linked) decision problems for all the Trusts.
Within this game formulation we can formally define the Nash equilibrium (cf.~Section~\ref{sec:intro} or \cite{Neumann,ShohamBrown2008}) by:
\begin{quote}
    A strategy profile $\hat{\mathbf{a}}=\left[\hat{\mathbf{a}}_1,\dots,\hat{\mathbf{a}}_N\right]$ is a \textit{Nash equilibrium} for the game $\left\{ G_1,\dots,G_N\right\}$ if and only if for all players $n\in\mathcal{N}$ we have
	\begin{equation}
		u_n\left(s_n^0,\left[\hat{\mathbf{a}}_n,\hat{\mathbf{a}}_{-n}\right]\right) \geq u_n\left(s_n^0,\left[\mathbf{a}_n,\hat{\mathbf{a}}_{-n}\right]\right),\ \ \forall \mathbf{a}_n\in\mathcal{H}_n\left(s_n^0\right)\ .
	\end{equation}
\end{quote}

\section{Materials and methods}
\label{sec:res}

In this paper, we propose a game-theoretic approach for the supply of PPE to healthcare facilities. We consider the case study of England as it was one of the countries severely impacted by the COVID-19 pandemic and where shortages in medical resources, especially PPE, were reported \cite{Tabah2020}. The players in this game structure are assumed to report their demand and are committed to paying their respective invoices. The players are also assumed to have the financial independence that inspires their selfish behaviour. Although this game is intended for independent healthcare providers, such as NHS Trusts in England, only region-level COVID-19 data are publicly available. Therefore, in this experiment, we assume that each of the seven NHS England regions can act selfishly in  a way that represents the interests of its accommodated Trusts. We believe this has no effect on the insights drawn from this experiment since a region is merely a group of Trusts.

\subsubsection*{Demand profiles until 01 Aug}

\begin{figure}[h]
\centering
\includegraphics[clip=true,trim={37pt 35pt 10pt 85pt},width=0.5\textwidth]{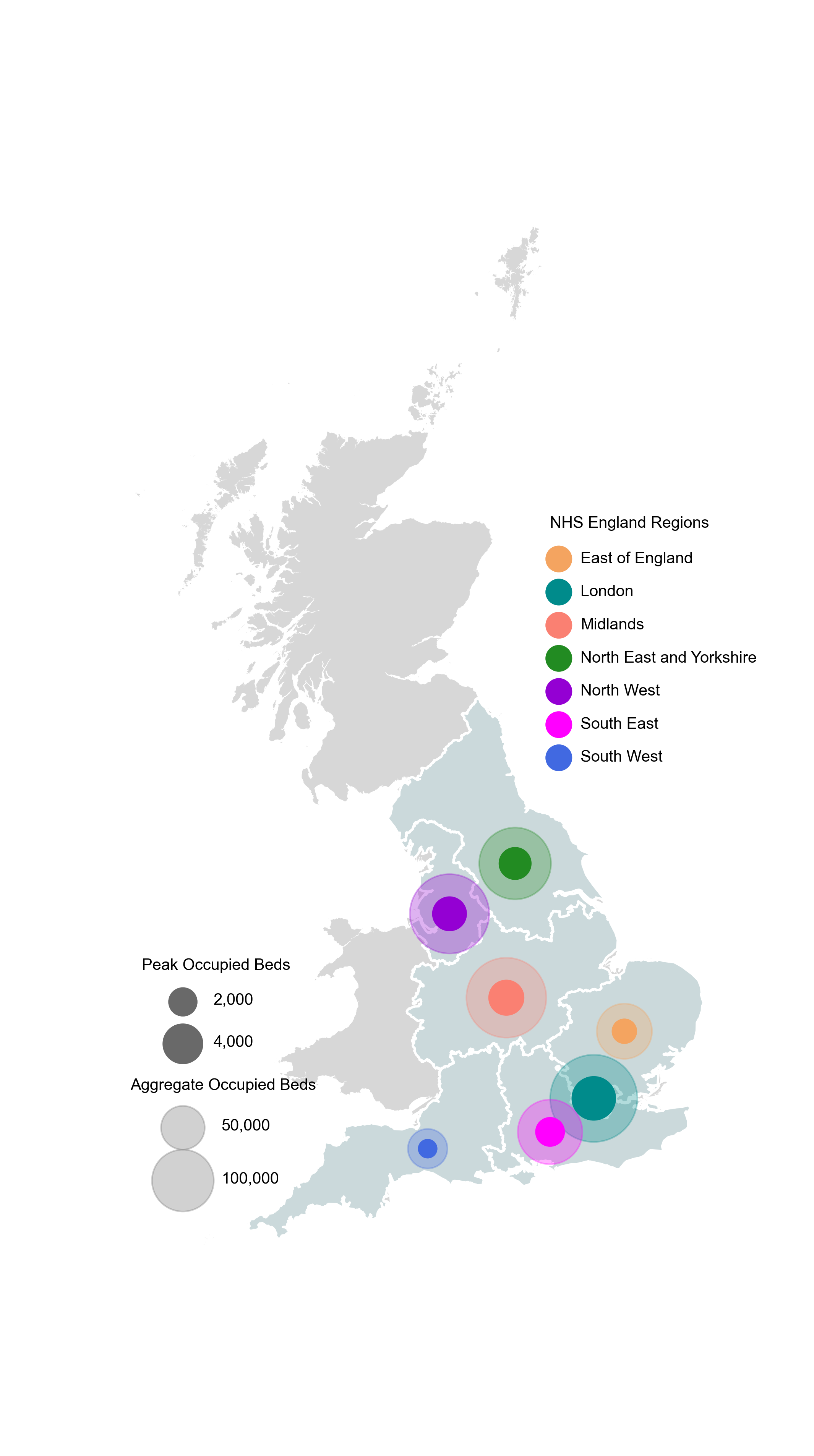}
\caption{\textbf{Total and peak occupied hospital beds with illness related to COVID-19 in the seven regions of NHS England up to and including 01 Aug 2020.} \footnotesize{Data to produce the background map in Fig~\ref{fig:Fig 2} is from the UK Office of National Statistics, 2019 licensed under the Open Government License \cite{OGLv32020}.}}
\label{fig:Fig 2}
\end{figure}

The region demand profiles used in this experiment were originated from daily COVID-19 occupied hospital beds data, which is available for the seven NHS England regions and published by the UK government \cite{PublicHealthEnglandPHE2020}. Fig~\ref{fig:Fig 2} shows the peak and total (up to 01 Aug) COVID-19 occupied hospital beds for each of the seven NHS England regions. We assume this information can represent the extent of the regions' response to the pandemic. Therefore, we assume that the PPE consumption that is related to all COVID-19 activities within these regions can be estimated from their daily COVID-19 occupied hospital beds data. As the collection of those daily data is already standard procedure in NHS hospitals, usage of the proposed model does not put any additional burden on NHS staff.

On 15 Apr, the UK government published its estimated PPE deliveries in England since the start of the outbreak in the country: “Since 25 February 2020, at least 654 million items of PPE have been supplied in this way” \cite{DepartmentofHealth&SocialCareDHSC2020}. As this number counts a pair of hand gloves as two items and includes consumables that are not considered for this study as mentioned in Section~\ref{sec:intro}, e.g.~body bags and swabs, we used the number of delivered aprons (135 M) as our reference for consumed PPE kits. However, only 86\% of these items were delivered to the NHS Trusts, whereas the rest were distributed to primary care providers (GPs), pharmacies, dentists and social care providers, as well as to other sectors \cite{DepartmentofHealth&SocialCareDHSC2020}. Additionally, although these deliveries took place during the period between 25 Feb to 15 Apr, the first COVID-19 hospitalization only took place on 20 Mar. Also, given that PPE supply chains were heavily burdened at the beginning of the outbreak, the UK government was led to change PPE usage guidance that health and social care workers should use in different settings when caring for people with COVID-19. The most notable of these was on 02 Apr when a single kit of PPE was advised to be used per session rather than per patient \cite{NationalAuditOfficeNAO2020}. 
This change of guidance along with the fact that our model is only concerned with COVID-19 related PPE consumption have led us to assume that 50\% of the PPE kits delivered to NHS Trusts between 25 Feb to 15 Apr were consumed for COVID-19 related activities in the period between 20 Mar and 15 Apr. This results in an estimated PPE consumption of 58 M kits for COVID-19 related hospitalisation cases. Dividing this number by the aggregated daily COVID-19 occupied beds in England for the aforementioned period results in a rough estimate of 195 PPE kits daily consumption per occupied hospital bed. 
Taking this into consideration, along with the mentioned change in guidance, we were able to extrapolate the publicly available daily COVID-19 occupied beds region data into PPE consumption by using a factor randomised between 210 and 240 kits per bed until 02 Apr, and between 150 and 180 from then onwards.

\subsubsection*{Storage Capacity}

According to the experience of our physician co-authors who have worked in several NHS hospitals, it can be estimated that each set of 20 hospital beds is supported by a PPE storage space of three shelves, the dimensions of each are $4.0\,m \times 0.6\,m \times 0.8\,m$. Our extrapolation in m\textsuperscript{3} to the storage capacity of the seven NHS England regions is listed in Table~\ref{Table 2}. Moreover, using the volumes supplied by PPE suppliers for bulk orders {\cite{Amazon.com, Amazon.coma, Amazon.comb, Amazon.comc}}, the volume of a thousand of PPE kits, each consisting of a face shield (visor), a face mask, an apron and a pair of gloves, can be approximated to 1.35 m\textsuperscript{3}. Consequently, the storage capacities in terms of PPE kits can be estimated for each region as seen in Table~\ref{Table 2}.

Although we are aware that our estimates are very coarse, we believe that they represent a reasonable attempt at quantifying those largely unpublished quantities. In any case, they are only used as a baseline for our modelling as four other values of storage capacities are also considered for each region, i.e. the baseline storage multiplied by a factor of 5, 10, 15 and 20.

\setcounter{table}{1}
\begin{table}[!ht]
\centering
\caption{\textbf{Storage capacities of the seven NHS England regions.}}
\label{Table 2}
\begin{tabular}{|p{4cm}|p{4cm}|p{1.5cm}|p{2cm}|}
\hline
Region & Maximum COVID-19 occupied beds & Storage in m\textsuperscript{3} & Storage in PPE kits\\ \thickhline
East of England & $1,484$ & $427$ & $316,590$\\ \hline
London & $4,813$ & $1,386$ & $1,026,770$\\ \hline
Midlands & $3,101$ & $893$ & $661,550$\\ \hline
North East \& Yorkshire & $2,567$ & $739$ & $547,630$\\ \hline
North West & $2,890$ & $832$ & $616,530$\\ \hline
South East & $2,073$ & $597$ & $442,240$\\ \hline
South West & $840$ & $242$ & $179,200$\\ \hline
\end{tabular}
\end{table}

\subsubsection*{Second wave}

Given that several countries have already experienced a second wave of COVID-19 infections (e.g.~Spain and France) \cite{Worldometer2020} and several studies \cite{Keeling2020, Panovska-Griffiths2020, Wise2020} predict that this is most likely to occur in England as well, our model is used to investigate how PPE demand should be handled in such a situation. Our experiments are informed by the model proposed  by \cite{Keeling2020},
where the authors predict that a second surge of COVID-19 cases happening in England can be of a lesser magnitude than the first wave, depending on the control measures in place (e.g.~lockdown and social distancing measures), but will most likely last longer than the first wave.

Therefore, a second surge in PPE demand was introduced to the regions' demand profiles from 01 Aug. Inspired by \cite{Keeling2020}, we modelled the second wave with a lower peak and taking place over a longer period. The second wave PPE demand data does not reflect results from a prediction model built for this study, but is merely a useful depiction for evaluating the performance of our model in case a putative second wave occurs. Fig~\ref{fig:Fig 3} shows an example of a second wave PPE demand profile expected to peak in mid October 2020. As the time of occurrence of a putative second wave is largely unknown, we considered in our experiments waves peaking at five different dates, i.e. mid October, mid November, mid December 2020 and mid January and mid February 2021. As can be shown from \cite{Keeling2020}, it is highly unlikely that a putative second wave would last longer than 100 days after peaking. This has led us to assume that PPE demand for the second wave would come to an end 100 days after the second peak.

\subsubsection*{Stockpiling starting dates}

In this study, we investigate whether shortages in PPE could have been avoided if the stockpiling game had been initiated at an earlier stage. Five different starting dates are considered as each of these had some importance regarding the development of the COVID-19 pandemic in the UK. Consequently, each could have triggered the start of the PPE stockpiling process:

\begin{enumerate}[itemsep=0pt, leftmargin=1cm, labelsep=0.3cm]
    \item 20 March 2020: data on COVID-19 cases and occupied hospital beds in England start to be released to the public \cite{PublicHealthEnglandPHE2020}. 
    \item 11 March 2020: the WHO declared COVID-19 a pandemic \cite{WorldHealthOrganizationWHO2020a}.
    \item 28 February 2020: the European Union proposed to the UK a scheme to bulk-buy PPE \cite{Boffey2020}.
    \item 07 February 2020: WHO warned of PPE shortages \cite{WorldHealthOrganizationWHO2020}.
    \item 31 January 2020: the first COVID-19 case was confirmed in the UK \cite{Bowden2020}.
\end{enumerate}

\subsubsection*{Cost parameters}
As mentioned in Section~\ref{sec:mat}, considerable rises in PPE costs were reported during peak COVID-19 cases in England \cite{Hardy2020}. This led us to use a quadratic cost function in our model to express cost as a function of demand (see Eq~\ref{eqn:quadratic_cost}). Since our proposed cost function does not have any cost terms other than the cost of ordering PPE, scheduling decisions are fairly independent of the cost parameters. In fact, the authors in \cite{pilz2019dynamic} showed that the sensitivity of similar scheduling games in relation to cost parameters is quite low. In the experiments conducted in this study, we use $8\times10^{-6}$ as the quadratic term parameter and $1\times10^{-2}$ as the linear term parameter. Usage of these parameters has resulted in a 300\% increase in PPE cost between peak and average demand. We believe this is in line with what has been reported in the literature \cite{Hardy2020}, where the costs of certain PPE items increased by up to 1000\% between pre-COVID and peak COVID periods.

\subsubsection*{Experiment implementation}

In order to show the outcome of the game and whether it has the capability to ease the challenge of securing PPE during peak demand, this experiment was implemented in two stages. 
In the first stage, we considered the overall period using five different stockpiling start dates (as stated above), to the end of a putative second wave, which also has five different peak dates. Additionally, we used five different storage capacities for each region in order to analyse the effects of adding extra PPE storage space. This means that 125 games were played in the first stage, the results of which will be discussed in the following sections. 
In the second stage of this experiment, we only examined the effects of running the game for the duration of the second wave; with the simulation starting from 01 Aug, (i.e.~the cut-off date of the data used to simulate demand), and ending 100 days after the peak of the second wave. We have simulated five different dates for the peak of the second wave.
Furthermore, we used the same five storage capacities for this stage as already employed before. The results of the 25 games that were simulated in the second stage will also be discussed in the following sections. In both stages, we assume that regions start the scheduling game with an empty store. We should specify that, since NHS hospitals received daily PPE orders during the considered period, the game was performed sequentially using a 24-hour interval to provide automated daily decisions fulfilling the PPE requirements of each Trust.

A Python code was written to find the Nash equilibrium of this game, where players sequentially update their PPE orders to minimise their costs. The optimisation package \textit{scipy.optimize} was used for each player to find their optimal scheduling decisions at each game iteration. The game converges once the mean squared error (MSE) between successive game iterations drop below a threshold. Given that peak PPE demand is in the order of millions of sets, we used a maximum value of 10 for the MSE to obtain accurate results.

\section{Results}
\label{sec:res2}

\subsection*{Analysis of outputs generated by the game}

\begin{figure}[h]
\centering
\includegraphics[width=\textwidth]{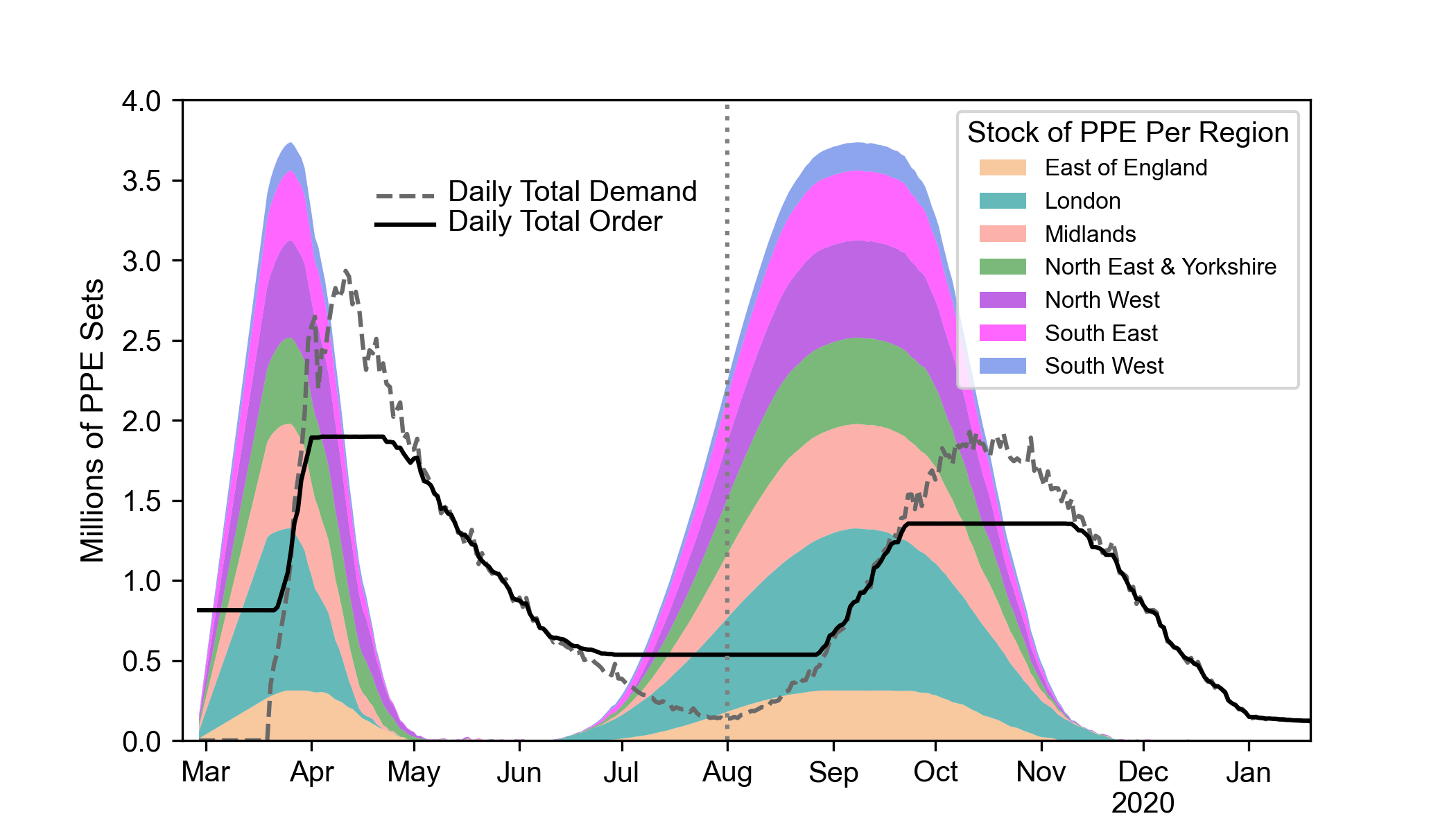}
\caption{\textbf{Millions of PPE sets required daily nationally (dotted line), ordered daily nationally (bold line) according to the game, and stored in each of the seven regions of NHS England (coloured areas).} Here, stockpiling started on 28 Feb, standard storage capacity was multiplied by five and the peak of the second wave is expected to happen in mid October. Stored PPE sets for each region is stacked on top of each other in order to show the cumulative PPE stock of all regions to illustrate the game mechanism. Note that the areas are scaled down by a factor of five in order to make the figure more readable.}
\label{fig:Fig 3}
\end{figure}

In order to illustrate the outputs generated by the game, we show results from running one of the games in Fig~\ref{fig:Fig 3}. 
The outcome of the game is a series of daily decisions that each English NHS region should take regarding their PPE requirements: order PPE, use PPE from their storage or stockpile PPE in their storage. The result of those decisions can be visualised by monitoring the status of their storage levels. Fig~\ref{fig:Fig 3} displays the amount of PPE sets available in the storage for each region (coloured areas), the PPE demand (dotted line) and PPE orders (bold line). From the stockpiling date, a constant set of PPE is ordered daily, leading to storage reaching full capacity at the start of the first wave that triggers PPE demand. However, as storage capacity becomes rapidly insufficient to meet the daily demand, available (stored) PPE decreases rapidly and daily ordering increases until reaching a plateau around the period of the peak of the first wave. Eventually, storage becomes depleted and the daily order equals the daily demand of a receding wave. Finally, when the daily demand is sufficiently low, storage is repleted allowing English regions to face the putative second wave with full PPE storage capacity. From then, patterns observed in terms of order and storage levels are similar to that already described for the first wave.

\begin{figure}[h]
\centering
\includegraphics[width=\textwidth]{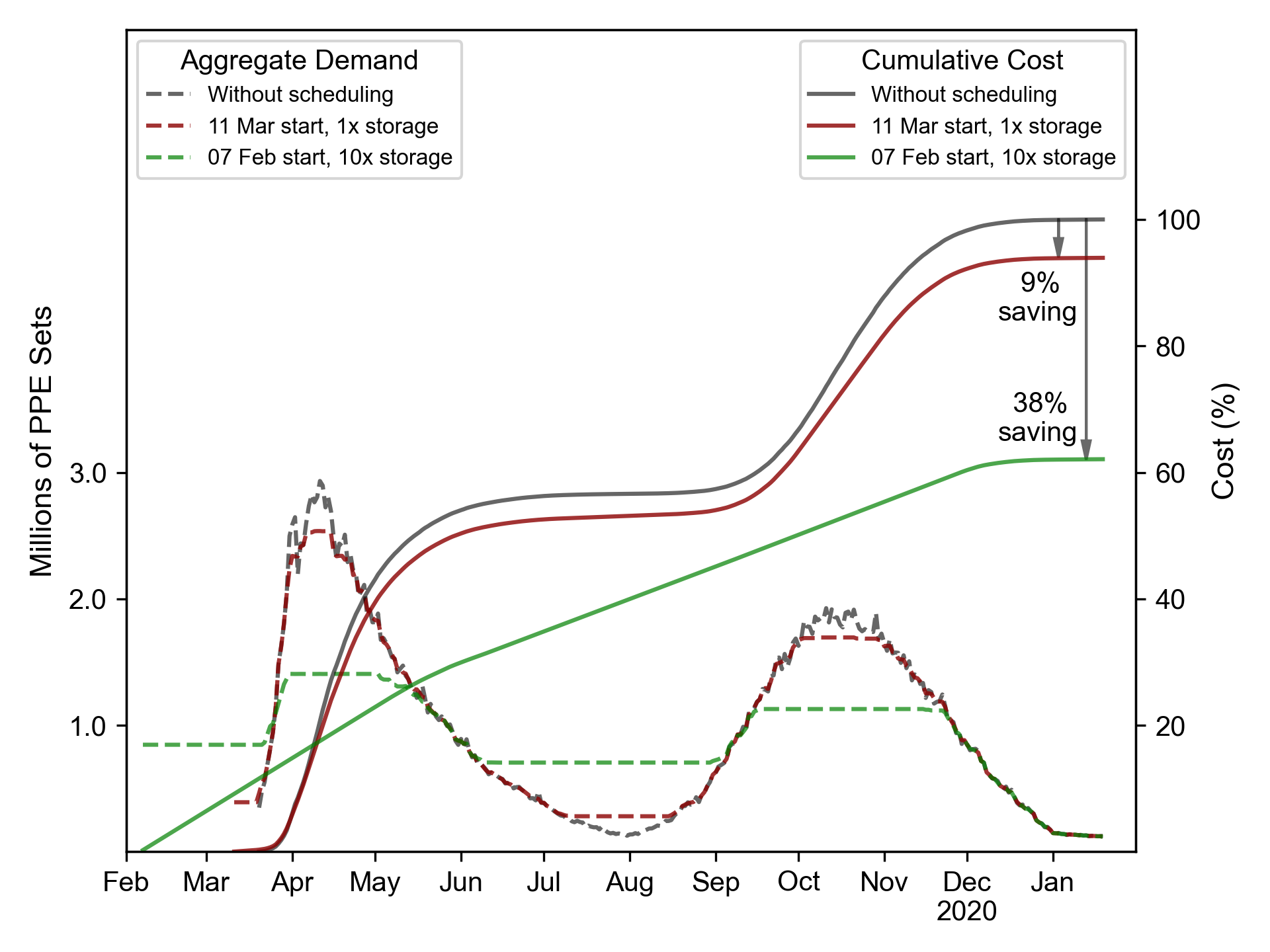}
\caption{\textbf{Comparison between the outcome of three different scenarios, a reference scenario (without scheduling), when stockpiling starts on 11 Mar and when it starts on 07 Feb and storage capacities are multiplied by 10.} This shows that cumulative cost saving can represent the challenge level of securing PPE supply. A colour grading is applied to represent the level of this challenge, where dark red is used for the highest challenge level and green for the lowest.}
\label{fig:Fig 4}
\end{figure}

As this model assumes that the cost of ordering PPE is quadratic with respect to the aggregated PPE order, the game aims at producing a level of orders as constant and low as the conditions permit. Deviations from this ideal scenario generate additional costs which corresponds in the proposed model to an increased challenge for the NHS at delivering required PPE. This challenge is visible in Fig~\ref{fig:Fig 4} where a comparison between three different scenarios is shown. The first scenario can be considered as a reference scenario where NHS regions report their PPE demand to NHS without scheduling their storage usage. The second and third scenarios are outcomes from the game when regions start stockpiling on 11 Mar and 07 Feb, respectively. One should also note that storage capacity in the third scenario is multiplied by 10. In this figure, the cumulative cost for the second and third scenarios is shown with reference to the first. As shown, a 9\% saving resulted from the second scenario at the end of the scheduling period while the third scenario resulted in a 38\% saving. Since a higher cost saving means that less fluctuations occur in demand, these figures can be directly used as a method to quantify the challenge of securing PPE. In Fig~\ref{fig:Fig 4}, as in the remaining figures, colours have been used to illustrate that level of challenge. The associated colour grading goes from dark red to dark green, where dark red expresses a high challenge and dark green indicates a relatively low challenge.

\subsection*{Scenario simulations}

\begin{figure}[h]
\centering
\includegraphics[width=\textwidth]{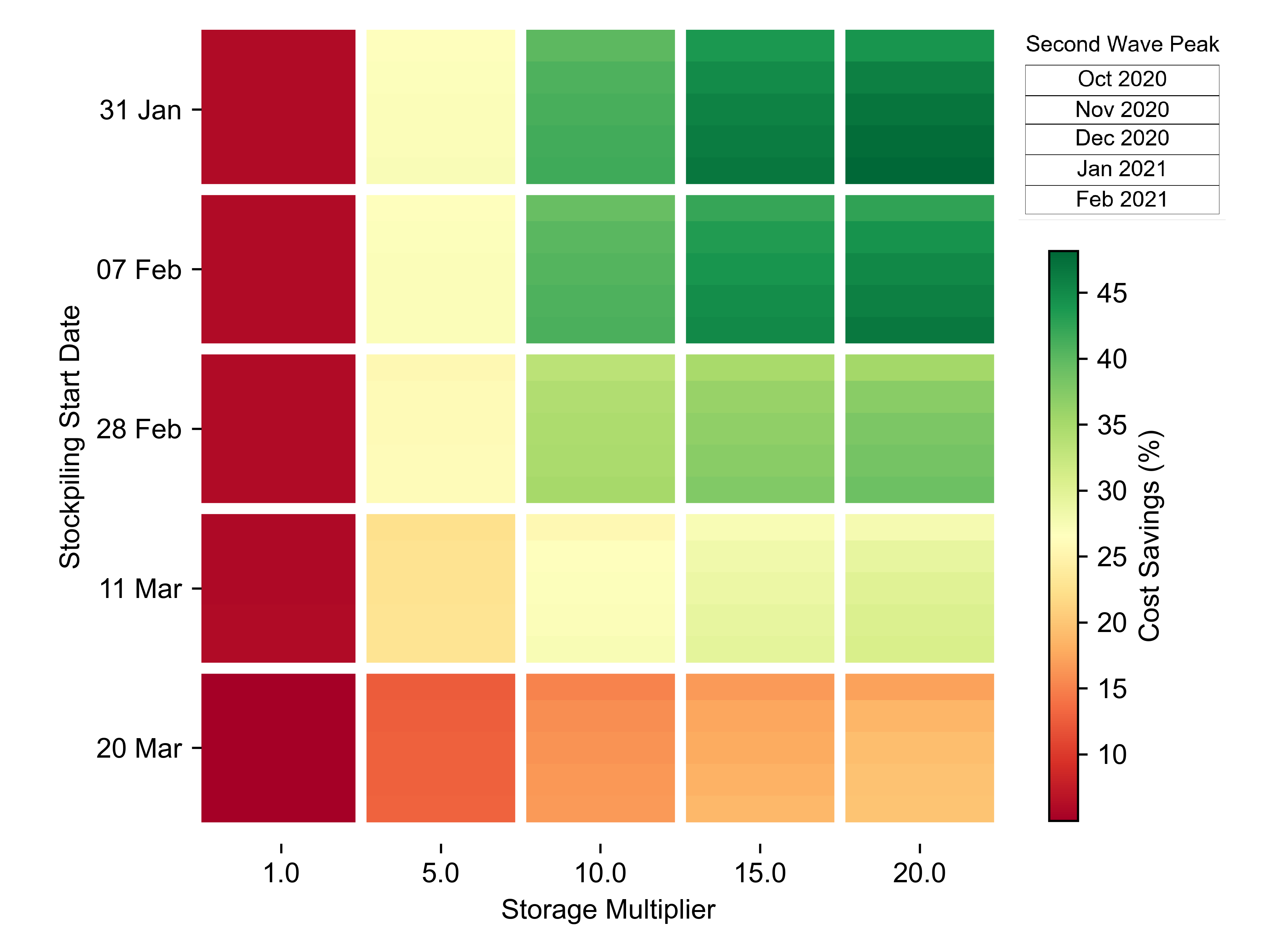}
\caption{\textbf{Challenge in terms of fulfilling PPE demand delivery according to the amount of available storage capacity (x-axis), the stockpiling starting date in 2020 (y-axis), and the peak date of a putative second wave of COVID-19} (each cell in the grid is divided in five stripes corresponding, from top to bottom, to peak dates in October, November, December 2020, January and February 2021).}
\label{fig:Fig 5}
\end{figure}

Using the proposed model, a range of different scenarios were considered to assess how the challenge in terms of fulfilling PPE demand varies according to the amount of available storage capacity, the stockpiling starting date and the peak date of a putative second wave of COVID-19. In total, 125 scenarios were conducted by considering five different values for each of the three parameters (see subsections above). Fig~\ref{fig:Fig 5} summarises the outcomes of these experiments.  
The figure reveals that in a two-wave pandemic, the two most critical parameters are the stockpiling date and storage capacity. With the peak of the first wave being estimated at 08 Apr \cite{Leon2020.04.21.20073049}, and the second wave taking place at least six months after the first one, the timing of the second wave thereafter would have only a moderate impact on easing the PPE challenge. 
On one hand, Fig~\ref{fig:Fig 5} highlights that if the storage capacity is low, an early stockpiling starting date hardly alleviates the PPE challenge and has minimal, if not negligible, cost savings (leftmost column). On the other hand, a high storage capacity only slightly mitigates a late stockpiling date (bottom row). Indeed, only early stockpiling and a sizeable increase in storage (top right area) would provide the right conditions to lower the PPE challenge overall and indeed would result in considerable cost savings.

\begin{figure}[h]
\centering
\includegraphics[width=\textwidth]{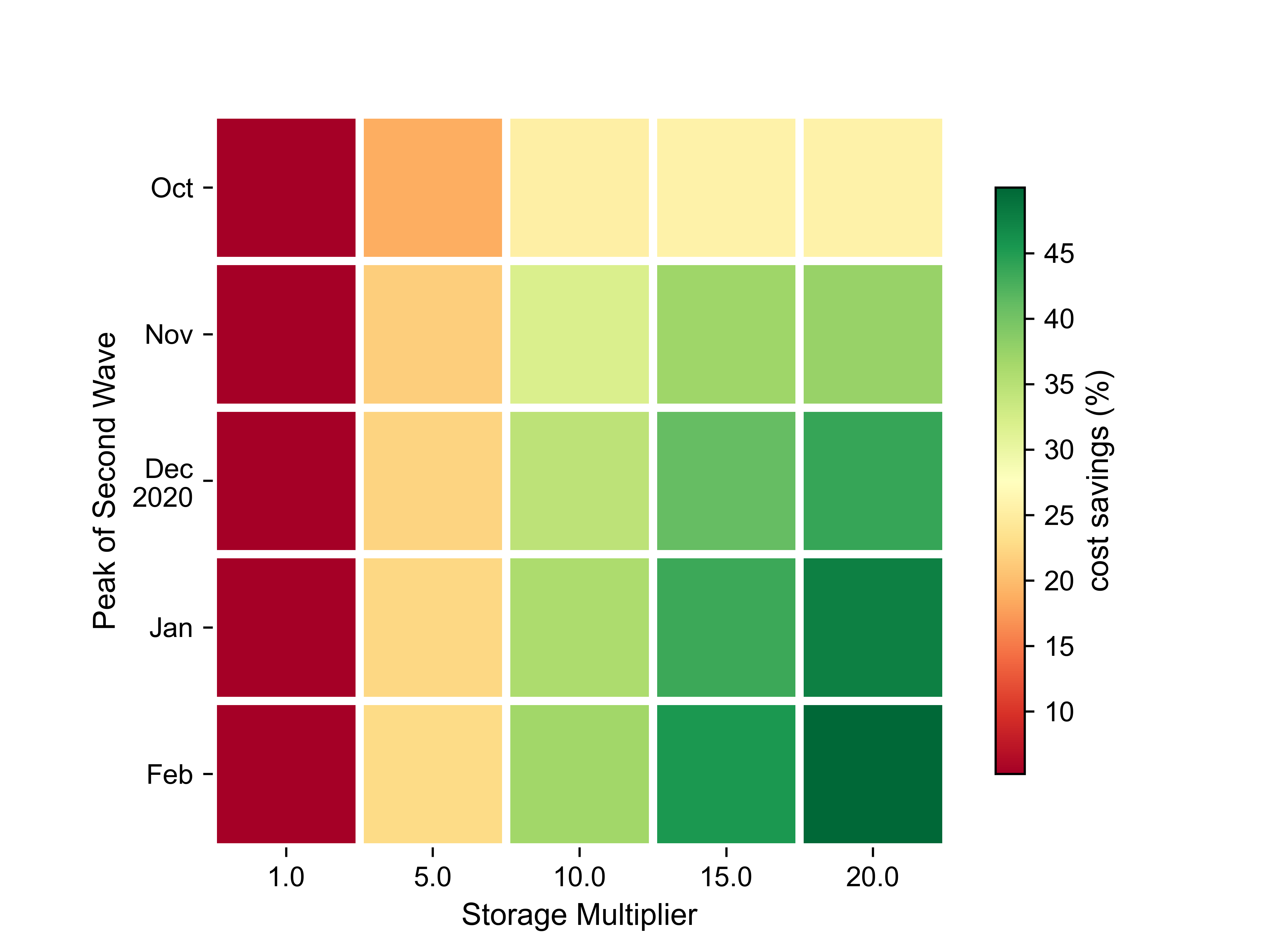}
\caption{\textbf{Challenge in terms of fulfilling PPE demand delivery according to the amount of available storage capacity (x-axis) and the peak date of a putative second wave of COVID-19 (y-axis).}}
\label{fig:Fig 6}
\end{figure}

In order to focus on the future, an additional 25 scenarios were simulated focusing on the PPE challenges associated with the available storage capacity and the peak date of a putative second wave. Using the last date of bed occupancy data (01 Aug) as the simulation starting date, five different values were considered for the two parameters (see subsections above). The results of these 25 scenarios are shown in Fig~\ref{fig:Fig 6}. As shown, while the storage capacity remains a key parameter, the impact of the date of the peak of the second wave has become apparent. Indeed, as the period of study is shorter than in the previous set of simulations and only a single wave of infection is considered, its timing substantially affects the PPE challenge. Consequently, even if a large amount of storage is available (two rightmost columns), only a late 2020 peak or later would be handled well in terms of cost savings.

\section{Discussion and recommendations}
\label{sec:dis}

The model we have constructed to analyse the provision  of PPE has important implications for the management of the pandemic hitherto. Firstly, we have shown objectively that early stockpiling and increasing storage capacity would have helped to massively reduce the costs associated with the containment of the pandemic. Secondly, the early, sufficient and cheap provision of PPE would have had a substantial impact on the ability to contain the pandemic, and protect both the most vulnerable patients and  the healthcare professionals that care for them \cite{EuropeanCentreforDiseasePreventionandControlECDC2020}.

Based on these findings and the mathematics underlying our approach, we have identified the key elements that govern the aggregated demand (and therefore, the cost) of PPE in our scenario. The two key parameters involved in saving costs by reducing the demand for PPE in the context of a pandemic that were identified using this approach are: (i) the storage capacity (for PPE) available to the health system. We have shown that increasing the storage capacity enhances the ability to stockpile which in turn helps flatten the demand curve (making it easier for suppliers to meet this demand in a cost-effective way); and (ii) the date when stockpiling of PPE commences. We have shown that earlier stockpiling would have saved costs and ensured adequate provision of PPE for the first peak of the pandemic. These two factors will also prove crucial when planning for a second wave. Of secondary importance is the ability to predict the date of the second wave. This is because, providing the second wave occurs after the new year, the degree of cost saving that occurs diminishes to negligible levels beyond this point in time.
Using our game-theoretic approach, we  have found that the NHS would be able to achieve all the cost-saving advantages if the second wave was to occur in 2021. A practical implication of this is that policymakers may need to keep some of the restricting measures that are in place to control the pandemic until beyond this point in time.

\begin{figure}[ht]
\centering
\includegraphics[width=\textwidth]{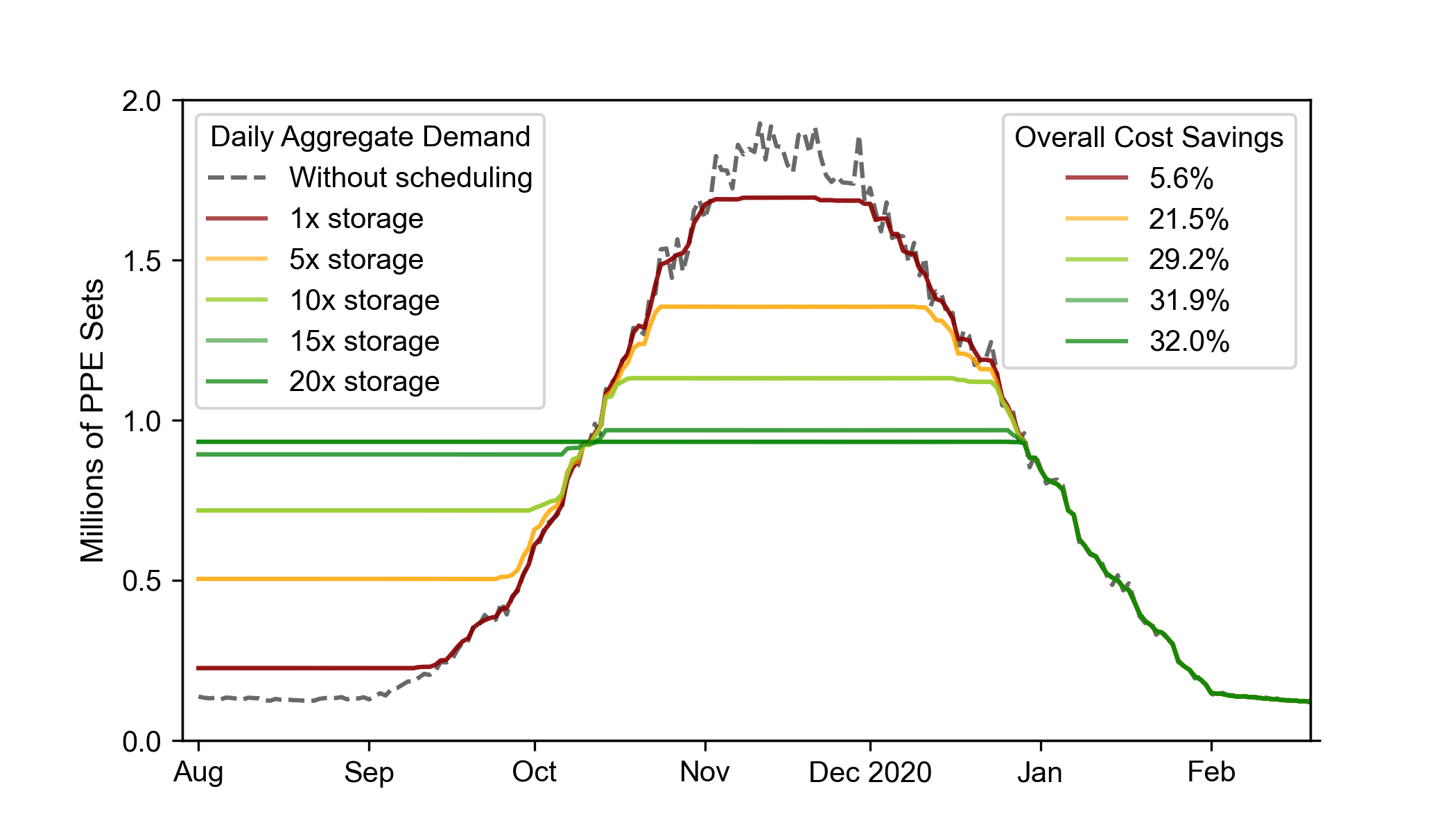}
\caption{\textbf{Effect of increasing storage capacity on the outcome of the game in terms of aggregate PPE demand.} These results are for a second wave where PPE demand peaks in mid November.}
\label{fig:Fig 7}
\end{figure}

In order to highlight the significance of having dedicated and sufficient storage for PPE in preparation for a putative second wave, we show a comparison between different storage capacities in Fig~\ref{fig:Fig 7}. In this figure, the aggregate demand profile for all NHS England regions is shown both with and without storage scheduling for a second wave of PPE demand that peaks in mid November. As shown, there is a substantial improvement in cost savings between the different storage capacities, especially when it is amplified by a factor of five and 10, resulting in an improvement in cost saving of 15.9\% and 7.7\% respectively. This means that the challenge of securing PPE during a second wave can be eased by a considerable amount when additional storage space is provided for stockpiling PPE. The figure also shows that further storage enhancement beyond this would result in less cost-saving improvements and would therefore have little impact on the challenge of securing sufficient PPE. This is especially visible when storage is expanded from 15 times to 20 times the standard storage space. Indeed, enhancing the PPE storage capacity by 15 times, perhaps by using temporary storage facilities as an interim solution, results in a relatively steady demand profile as shown in Fig~\ref{fig:Fig 7}.

\section{Conclusion}
\label{sec:con}
In this paper, we developed a game-theoretic model for scheduling PPE supply for healthcare facilities. In this discrete time dynamic game, healthcare facilities make stockpiling decisions that minimise their PPE ordering cost. Our model adopts a centralised-decentralised approach to the PPE supply chain, where a central entity controls the PPE pricing formula, yet is committed to fulfilling the orders independently placed by healthcare providers. Based on publicly available COVID-19 hospitalisation data for NHS England regions, we performed simulations to investigate the impact of three key factors on PPE security of supply. These factors comprise: (i) the stockpiling start date, (ii) the time of the peak of a putative second wave, and (iii) the amount of storage available for medical PPE. The two most critical parameters were found to be the storage capacity and the stockpiling start date, while the timing of the second wave has only had a moderate impact on the challenge of securing PPE. Within our model we observe that enhancing PPE storage capacities by a factor of 15 is sufficient to considerably lower the peak demand at any given day and effectively minimises the strain on the health care system. (This does not imply that dealing with the pandemic in this scenario is rendered trivial. It rather represents the most favourable way to act during the crisis).

While the shortage of PPE supplies for care home workers during the first peak in the UK was a topic of great concern, since those shortages were blamed for care home outbreaks which led to a large number of lost lives \cite{Brainard2020}, we explicitly excluded this from our study due to the lack of publicly available data. Access to such data would provide further insights into demand patterns which can potentially influence the game. Extensions to our work can include introducing supply constraints where demand cannot always be fulfilled, perhaps by adding a shortage penalty to the cost function. Also, our model assumes that healthcare facilities can predict their demand without forecasting errors. This can be enhanced by adding stochastic elements that capture the uncertainty in demand \cite{pilz2019dynamic}. The model can also be extended by investigating scenarios where hospitals can share their resources by having mutual agreements or using the selfish sharing approach proposed in \cite{Pilz2019SmartGridComm}. This could be via using a scheme to incentivise matching scheduled regional demand with the production of the supplier. Another possible extension of this research is to propose a model that finds the optimal storage capacity for each game player. 
Although the above refinements may give greater insight, our model already has the potential to be a useful tool that can help governments and policy makers in decision-making in relation to critical PPE supplies. Finally, if one wishes to extend this work to incorporate the global picture, this could be achieved by introducing another layer of competition modelling the orders of the different nations. Here, a Leader-Follower structure would  be required and most suitably captured by a Stackelberg game \cite{ShohamBrown2008}.

\section*{Acknowledgments}

The authors dedicate this work to healthcare workers that have sadly lost their lives during this pandemic, and would like to thank all of those working on the front-lines, whether it is in hospital wards, care homes or research laboratories.

%
%
%

\end{document}